\documentclass{ws-ijmpcs}

\begin{document}

\markboth{R. Valentim, J. E. Horvath \& E. M. Rangel}
{Insert running title (not more than 8 words)}

%
%

\title{Evidence for two neutrinos bursts from SN1987A}

\author{R. Valentim}

\address{Departamento de F\'{\i}sica, ICAFQ, Universidade Federal de S\~ao Paulo, \\
Diadema, S\~ao Paulo 09913-030 ,
Brazil\\
valentim.rodolfo@unifesp.br, r.valentim@gmail.com}

\author{J. E. Horvath}

\address{Departamento de Astronomia, IAG, Universidade de S\~ao Paulo, R. do Mat\~ao, 1226 - Butant\~a, S\~ao Paulo - SP, 05508-090, Brazil\\
foton@iag.usp.br}

\author{E. M. Rangel}

\address{Engebr\'as\\
rangelfisica@gmail.com}

\maketitle


\begin{abstract}

The SN1987A in the Giant Magellanic Cloud was an amazing and extraordinary event because it was detected in real time for different neutrinos experiments ($\nu$s) around the world. Approximate $\sim25$ events  were observed in three different experiments:  Kamiokande II (KII) $\sim 12$, Irvine-Michigan-Brookhaven (IMB) $\sim 8$ e Baksan $\sim 5$, plus a contrived burst at Mont Blanc (Liquid Scintillator Detector - LSD) later dismissed because of energetic requirements (Aglietta et al. 1988). The neutrinos have  an important play role into the neutron star newborn:  at the moment when the supernova explodes the compact object remnant is freezing  by  neutrinos ($\sim99\%$ energy is lost in the few seconds of the explosion). The work is motivated by neutrinos' event in relation arrival times where there is a temporal gap between set of events ($\sim6\mbox{s}$). The first part of dataset came from the ordinary mechanism of freezing and the second part suggests different mechanism of neutrinos production.  We tested two models of cooling for neutrinos from SN1987A: 1st an exponential cooling is an ordinary model of cooling and 2nd a two-step temperature model that it considers two bursts separated with temporal gap.  Our analysis was done with Bayesian tools ({\it Bayesian Information Criterion} - BIC) The result showed strong evidence in favor of a two-step model against one single exponential cooling ($\ln\mbox{B}_{ij} > 5.0$), and suggests the existence of two neutrino bursts at the moment the neutron star was born.

\end{abstract}

\ccode{PACS numbers:}

\section{Introduction}	

The SN1987A explosion was the first amazing event observed by modern neutrinos detectors. Few hours after the explosion neutrinos experiments around the world  (Bionta 1987, Hirata et. al. e  Alekseev et. al 1987)  observed neutrinos burst associated to gravitational collapse of supernova. After the neutronization, electrons combined with protons through beta inverse decay ($\mbox{p}^{+}+\mbox{e}^{-}\rightarrow \nu+\mbox{n}$) and the compression of the stone increases until it becomes very rigid. With the advent of collapse, the outermost layers of the star fall on the core suffering a rebound effect, it is important to note that the whole process is not completely understood. After the collapse the supernova explodes causing the complete ejection of the envelope. Moments before the explosion, the iron core that had $\simeq500\mbox{km}$ of radius is compressed into a sphere of $30-40\mbox{km}$ containing an enormous amount of thermal energy product of the implosion. Before obtaining a stable configuration, the newly born neutron star irradiates neutrinos through processes such as bremsstrahlung ($\mbox{n}+\mbox{n}\rightarrow\mbox{n}+\mbox{n}+\nu+\bar \nu$) of neutrons and annihilation of pairs ($\mbox{e}^{+}+\mbox{e}^{-}\rightarrow \nu + \bar \nu$) (ref). These neutrinos come from the burst and carry $99\%$ of the event's energy in the cooling process that forms the neutron star remnant. The emission spectrum is like a black body of fermions (Nadyozhin $\&$ Otroshenko 1980). The neutrinos luminosity is $\mbox{L}_{\nu} \propto \mbox{R}^2_{\nu}\mbox{T}^4_{\nu}$ where $\mbox{R}_\nu$ is the neutrinosphere radius and $\mbox{T}_\nu$ is the temperature of the neutrinosphere and is calculated from the mean energy of the observed neutrinos and  averaged energy with Fermi function of the source, the estimated value is $\propto\mbox{T}_{\nu}= 4.2^{+1.2}_{-0.8}\mbox{MeV}$.  The data observed by KII, IMB and Baskan ($\sim25$ events) point to a temporal distribution with a gap of $\sim5$ seconds between the observed neutrinos (referencia). The temporal evolution of temperature showed that the possibility the some hiatos (Loredo $\&$ Lamb 2002) and this suggests the idea of ??a second burst was proposed by Benvenuto $\&$ Horvath (1989). The authors proposed the second burst as coming from Strange Quark Matter (SQM) scenarios.

\section{Statistical Methodology}

Statistical tools used in this work was Bayesian Inference this approach allows compare two models with different parameters numbers (Loredo $\&$ Lamb 2002). Bayesian Information Criterion (BIC) tests and compares two models from likelihood function that was built by Loredo $\&$ Lamb (2002). Likelihood function considers neutrinos emission from neutron star remnant, propagation of signal and detection at the Earth. The emission was modeled as Fermi's black-body (Nadyozhin $\&$ Otroshenko 1980), when the remnant was formed before the explosion the cooling mechanism emits neutrinos as black-body, the temperature of neutriosphere is on Fermi function. The propagation of signal considers the neutrino number flux per unit energy incident on detectors at the Earth. The detection envolves two distinct processes: 1o. a neutrino somehow produce an energetic charged lepton in the detector. 2o. the Cherenkov light produced in the detector by this charged lepton must be detected. More details of this is Loredo $\&$ Lamb (2002).  The explicit likelihood function expression is:

 \begin{equation}
\begin{split}
\mathcal{L(\mathcal{P})}=exp\left[-f\int_T{dt}\int{d\vec{n}}\int{\bar{\eta}(\vec{n},\epsilon)}R(\vec{n},\epsilon,t_i)\right] \\
\times\prod^{Nd}_{i=1}e^{R_{eff}
(t_i)\tau}\left[B_i + \int{d\vec{n}}\int{d\epsilon\mathcal{L}_i(\vec{n},\epsilon)R(\vec{n},\epsilon,t_i)}\right].
\end{split}
\end{equation}

Where $\mbox{f}$ is the fraction of the detector duty cycle, $\tau$ is dead time, $\bar{\eta}(\vec{n},\epsilon)$ efficiency curve of the detector, $B_i$ 
is the rate of noise integrated on time, $\mathcal{L}(\vec{n},\epsilon)$ weight function e $R(\vec{n},\epsilon,t_i$) rate measured events. Both models assumes the neutrinosphere radius is correlated with $\alpha$ that is a parametrization (Loredo $\&$ Lamb 2002). 

\subsection{Models}

The first model is a plain exponential cooling (EC), assuming just one continuous emission expected from a conventional physics assumption. Of course, the success 
of the explosion itself has to be assumed and is never tested as such. The evolution of the temperature is

\begin{equation}
\mbox{T}(t)= \mbox{T}_0\exp{(-t/4\tau)};
\end{equation}

this model has three parameters: $\mbox{T}$ (temperature), $\alpha$ time of decay and $\tau$ is time where the neutrinos' luminosity is constant. This model allows to test the hypothesis that the neutrino's flux decreases with time.

The second model, appropriated for the physical picture of Benvenuto \& Horvath (1989), is a Two-Step Temperature (TST):

\begin{equation}
\label{eq01}
T(t) = \left\{
\begin{array}{l l}
T  & \quad \mbox{para $0<t<1.0 s$} \\
0.1 MeV & \quad \mbox{para $1.0<t <1.0+\Delta t$} \\
a_p T &\quad\mbox{para $1.0+\Delta t<t<5.0+\Delta t$};

\end{array} \right.
\end{equation}

this model has also four parameters: $\mbox{T}$ (temperature), $\mbox{a}_p$ (scale factor), $\Delta t $(interval between the two emissions at different temperatures) and $\alpha$ (it considers the neutrinosphere radius constant at the emission moment). 
This second peak can be associated to a phase transition to strange matter, which proceeds extremely fast and resets the thermal content of the 
now proto-strange star (Benvenuto $\&$ Horvath 1989).

\subsection{Results}

The results are showed in the Table below. The best fits for the EC features $\mbox{T}=3.47\mbox{MeV}$, neutrino time on constant luminosity $\tau=4.75\mbox{s}$ 
and a neutrinosphere radius $\mbox{R}=34.7\mbox{km}$. The Two-Step Temperature (TST) suggests the following values for the temperatures, neutrinosphere radius and the temporal gap between bursts: 
$\mbox{T}_1 =3.91\mbox{MeV}$ and $\mbox{T}_2 = 2.81\mbox{MeV}$, a neutrinosphere radius of $\mbox{R}_{\nu}=31.60\mbox{km}$ (within the same range for both 
emissions), and a temporal gap $\Delta t = 5.40\mbox{s}$.

It is worthwhile to stress that when the Bayesian Information Criterion is employed to compare both models with different parameters, the TST model is 
always more likely than EC model $\ln{B}_{ij}>5.0$, this result suggests existence of two bursts and not just only one. 

\begin{table}[htb]
   \centering   
   \large       
   \setlength{\arrayrulewidth}{2\arrayrulewidth}  
   \setlength{\belowcaptionskip}{10pt}  
    \caption{Intervalos de confian\c{c}a frequentista dos par\^ametros para os modelos de resfriamento exponencial (EC) e duas temperaturas (Step Two Temperatures).}

   \begin{tabular}{cccccccc} 
      \hline\hline
      $Model$               & $\mbox{R}(10km)$      & $\mbox{T}(\mbox{MeV})$ & $\tau(\mbox{s})$ & $\Delta\mbox{t}(\mbox{s})$ & $\mbox{ap}$ \\
      \hline

       $EC$                    &  $3.73$     & $3.47$ & $4.75$ & $-$ &$-$   \\

       $TST$                    &  $3.16$     & $3.91$ & $-$  & $5.40$ & $0.72$ \\

       \hline\hline

    \end{tabular}
   \label{tab:Referencia_desejada}
   \end{table}

Further studies of this problem are guaranteed. In particular, it is relevant to study whether two exponential decays (instead of two constant temperatures) 
are still favored using BIC, since the number of parameters will be 5 and the formalism penalizes the growing number of them. 
 
\section*{Acknowledgments}

Both authors acknowledge research grants from FAPESP (Processes no. 2013/26258-4 and 2016/09831-0). J.E.H. has been 
financed by the CNPq Federal Agency through a Research Fellowship.

\section{References}


\end{document}